\documentclass[prb,aps,twocolumn,amsmath,amssymb,showpacs]{revtex4}
\usepackage{graphicx}
\usepackage{psfrag}
\usepackage{color}
\usepackage{subfigure}
\usepackage{amsmath}
\definecolor{dred}{rgb}{0.7,0.0,0.0}
\allowdisplaybreaks 
\usepackage{array}

\newcommand{\BE}{\begin{equation}}
\newcommand{\BEA}{\begin{eqnarray}}
\newcommand{\EE}{\end{equation}}
\newcommand{\EEA}{\end{eqnarray}}

\newcommand{\e}{{\rm e}}

\begin{document}

%
%

\title{Mott transition and ferrimagnetism in the Hubbard model 
on the anisotropic kagom\'e lattice}
\author{A. Yamada$^1$, K. Seki$^1$, R. Eder$^{2}$, and Y. Ohta$^1$}
\affiliation{$^1$Department of Physics, Chiba University, Chiba 263-8522, Japan\\
$^2$Karlsruhe Institut of Technology,
Institut f\"ur Festk\"orperphysik, 76021 Karlsruhe, Germany}

\date{\today}

\begin{abstract}

Mott transition and ferrimagnetism are studied in the Hubbard model on the anisotropic kagom\'e lattice 
using the variational cluster approximation and the phase diagram at zero temperature and half-filling 
is analyzed. 
The ferrimagnetic phase rapidly grows as the geometric frustration is relaxed, 
and the Mott insulator phase disappears in moderately frustrated region, 
showing that the ferrimagnetic fluctuations stemming from the relaxation of the geometric frustration is enhanced by the electron correlations. 
In metallic phase, heavy fermion behavior is observed and mass enhancement factor is computed. 
Enhancement of effective spatial anisotropy by the electron correlations 
is also confirmed in moderately frustrated region, and its effect on heavy fermion behavior is examined. 

\end{abstract}
 
 
\maketitle

%
%
Effect of geometric frustration is one of the most important subjects actively studied 
in the field of strongly correlated electron systems. 
For instance, 
the heavy fermion behavior in $\mathrm{LiV_2O_4}$ \cite{kondo97,jonsson07} with pyrochlore lattice structure, and 
the spin liquid states in the triangular-lattice organic materials $\kappa$-(BEDT-TTF)$_2\mathrm{X}$ 
\cite{lefebvre00,shimizu03,kagawa04} and herbertsmithite $\mathrm{ZnCu3(OH)_6Cl_2}$ 
with kagom\'e lattice structure \cite{helton07,mendels07} have attracted a lot of attentions. 

When spatial anisotropy is introduced in systems with geometric frustration, the interplay between the spin fluctuations 
and Mott transition appears as a new feature and provides unique phenomena which take place neither in the unfrustrated nor 
fully frustrated systems. A reentrant behavior of the Mott transition 
observed in the $\kappa$-(DEBT-TTY)$\mathrm{_2Cu[N(CM)_2]Cl}$ 
under pressure \cite{lefebvre00,kagawa04} is an interesting example realized on an anisotropic triangular-lattice, where 
that behavior stems from the enhancement of the 
antiferromagnetic fluctuations due to the electron correlations.\cite{ohashi08,liebsch09}

\begin{figure}
\includegraphics[width=0.47\textwidth,trim = 0 0 0 0,clip]{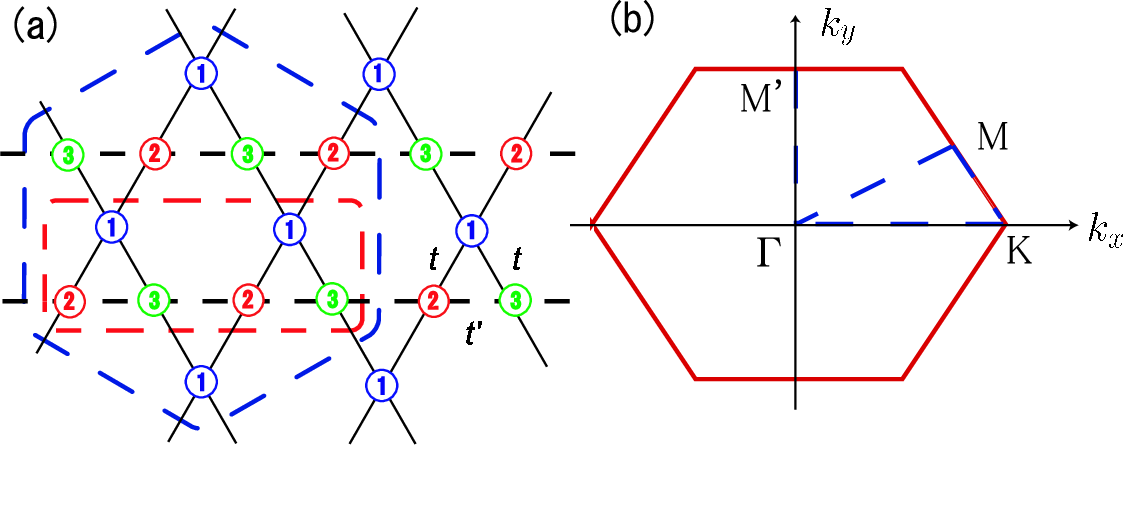}
\caption{(Color online) 
(a) Anisotropic kagom\'e lattice. In our lattice geometry, the three sites $1,2$, and $3$ form an equilateral triangle of the unit length 
and the dashed lines are along the $x$ direction. 
Inside the hexagon (dotted line) and square (dash-dotted line) are the 12- and 6-site clusters, respectively, 
which will be used in our analysis. 
(b) The first Brillouin zone of the anisotropic kagom\'e lattice. 
\label{fig:model}\\[-1.5em]}
\end{figure}

As for the kagom\'e lattice, which is a prototype of frustrated systems, a fully frustrated case has been theoretically studied 
in detail,\cite{imai03,bulut05,ohashi06,udagawa10} however the issues related to the anisotropy have been considered only recently. 
The Mott transition and magnetic properties near the transition have been studied using the cellular dynamical mean field theory\cite{kawakami10}, 
where the Mott transition point was analyzed and enhancement of spatial anisotropy and spin correlations were 
observed. 
Such enhancement may give rise to the extension of the ordered (ferrimagnetic) phase. 
Therefore, if the Mott transition itself persists without being veiled by the ferrimagnetic phase remains to be examined. 
Also, the effect of the enhanced anisotropy on the heavy fermion behavior is worth being studied. 

In this paper, we investigate the ferrimagnetism and Mott transition on the anisotropic 
kagom\'e lattice using the variational cluster approximation (VCA)\cite{Senechal00,Potthoff:2003-1,Potthoff:2003}, 
which is formulated based on a rigorous variational principle and exactly 
takes into account the short-range correlations. 
We study the phase diagram at zero temperature and half-filling. 
We show that, in moderately frustrated region, the ferrimagnetic phase rapidly grows down to the metal-insulator phase boundary, 
indicating that the spin correlations stemming from the relaxation of the frustration is enhanced by the electron correlations 
and the Mott insulator (MI) phase disappears. 
In the metallic phase, heavy fermion behavior is observed and the mass enhancement of the quasiparticle is computed. 
Effective spatial anisotropy becomes also larger due to the electron correlations, in agreement with 
the previous study.\cite{kawakami10} 
This effect gives rise to an enhancement of the anisotropy of the effective masses of the quasiparticles.

%
%

The Hamiltonian of the Hubbard model on the anisotropic kagom\'e lattice (see Fig.~\ref{fig:model}) reads 
\begin{align}
H =& -\sum_{i,j,\sigma} t_{ij}c_{i\sigma }^\dag c_{j\sigma}
+ U \sum_{i} n_{i\uparrow} n_{i\downarrow} - \mu \sum_{i,\sigma} n_{i\sigma},
\label{eqn:hm}
\end{align}
where $t_{ij}=t$ between the sites 1 and 2, 3 and $t_{ij}=t'$ between the sites 2 and 3, $U$ is the on-site Coulomb repulsion, and $\mu$ is the chemical potential. 
The annihilation (creation) operator for an electron at site $i$ with spin $\sigma$ is denoted as 
$c_{j\sigma}$ ($c_{i\sigma }^\dag$) and $n_{i\sigma}=c_{i\sigma}^\dag c_{i\sigma}$. 
The system corresponds to the fully frustrated kagom\'e lattice at $t'/t = 1$, and frustration becomes weaker with 
decreasing $t'/t$. The end member at $t'/t=0$ is a decorated square lattice. 
The energy unit is set as $t=1$ hereafter.

%
%

We use VCA\cite{Senechal00,Potthoff:2003-1,Potthoff:2003} to examine the phase diagram and 
behavior of the quasiparticles in the metallic phase at zero temperature. 
VCA is an extension of the cluster perturbation theory\cite{Senechal00} based on the 
self-energy-functional approach.\cite{Potthoff:2003} 
This approach uses the rigorous variational principle 
$\delta \Omega _{\mathbf{t}}[\Sigma ]/\delta \Sigma =0$ for the thermodynamic grand-potential 
$\Omega _{\mathbf{t}}$ written as a functional of the self-energy $\Sigma $
\begin{equation}
\Omega _{\mathbf{t}}[\Sigma ]=F[\Sigma ]+\mathrm{Tr}\ln
(-(G_0^{-1}-\Sigma )^{-1}).
\end{equation}%
In the above expression, $F[\Sigma ]$ is the Legendre 
transform of the Luttinger-Ward functional\cite{lw} and the index $\mathbf{t}$ denotes the explicit dependence of 
$\Omega _{\mathbf{t}}$ on all the one-body operators in the Hamiltonian. 
The stationary condition for $\Omega_{\mathbf{t}}[\Sigma ]$ leads to the Dyson's equation. 
All Hamiltonians with the same interaction part share the same functional form of $F[\Sigma ]$, and using that property 
$F[\Sigma ]$ can be evaluated from the exact solution of a simpler Hamiltonian $H'$, though 
the space of the self-energies where $F[\Sigma ]$ is evaluated is now restricted to that of $H'$. 
In VCA, one uses for $H'$ a Hamiltonian formed of clusters that are disconnected by removing hopping terms 
between identical clusters that tile the infinite lattice. 
A possible symmetry breaking is investigated by including 
in $H'$ the corresponding Weiss field that will be determined by minimizing the 
grand-potential $\Omega _{\mathbf{t}}$. 
Rewriting $F[\Sigma ]$ in terms of the grand-potential $\Omega'\equiv \Omega'_\mathbf{t}[\Sigma]$ and 
Green function $G'{}^{-1}\equiv G'_0{}^{-1}-\Sigma$ of the cluster Hamiltonian $H'$, the grand-potential is expressed as 
\begin{equation}
\Omega _{\mathbf{t}}(\mathbf{t}')=\Omega'\kern-0.4em - \kern-0.4em\int_C{\frac{%
d\omega }{2\pi }} \e^{ \delta \omega} \sum_{\mathbf{K}}\ln \det \left(
1+(G_0^{-1}\kern-0.2em -G_0'{}^{-1})G'\right)\label{omega}
\end{equation}%
and is now a function of $\mathbf{t}'$. 
The functional trace has become an integral over the diagonal variables 
(frequency and superlattice wave vectors) of the logarithm of a determinant over intra-cluster indices. 
The frequency integral 
is carried along the imaginary axis and $\delta \rightarrow + 0$. 
The stationary solution of $\Omega _{\mathbf{t}}(\mathbf{t}')$ and the 
exact self-energy of $H'$ at the stationary point, denoted as $\Sigma^{*}$, are the approximate grand-potential 
and self-energy of $H$ 
in VCA, and physical quantities, such as expectation values of the one-body operators, are calculated using 
the Green function $G_0{}^{-1}-\Sigma^{*} $. 
In VCA, the restriction of the space of the self-energies $\Sigma$ into that of $H'$ 
is the only approximation 
involved and short-range correlations within the cluster are exactly taken into account by exactly solving $H'$. 

In our analysis, the 6- and 12-site clusters in Fig.~\ref{fig:model}(a) are used to set up the cluster Hamiltonian $H'$. 
These clusters have even number of sites so that a singlet ground state is possible. To study the ferrimagnetism, the Weiss field 
\begin{eqnarray}
H_{\rm F}&=& h_{\rm F}\sum_{i} {\rm sign}(i)(n_{i\uparrow }-n_{i\downarrow }) 
\end{eqnarray}
with ${\rm sign}(i) = -1$ for the site 1 and ${\rm sign}(i) = 1$ for the sites 2 and 3, is also included. 
In the stationary point search of $\Omega(\mu', h_{\rm F})$, which we denote as the grand-potential per site, 
the Weiss field $h_{\rm F}$ and the cluster chemical potential $\mu'$ are treated as the variational parameters, where 
the latter should be included for the thermodynamic consistency.\cite{aichhorn} 
During the search, the chemical potential of the system $\mu$ is also adjusted so that the electron 
density $n$ is equal to 1 within 0.1\%. 
In general, a stationary solution with $h_{\rm F} \neq 0$ corresponding to the ferrimagnetic state and 
that with $h_{\rm F} = 0$ corresponding to the paramagnetic state are obtained, and the ground-state energies per site 
$E=\Omega+\mu n$ are compared to determine which solution (ferrimagnetic or paramagnetic) is stable. 
The density of state 
\begin{eqnarray}
\label{OPeq}
D(\omega)= \lim_{\eta \rightarrow 0}  \int
{\frac{%
d^2 k }{(2\pi)^2 }}\sum_{\sigma, a=1}^{3}\{ -\frac{1}{\pi} \mathrm{Im}G_{a\sigma}(k, \omega+i\eta) \}
\end{eqnarray}
is also calculated to examine the gap. 
To be precise, $D(\omega)$ is calculated for $\eta =0.2$, $0.1$, and $0.05$, and $\eta \rightarrow 0$ limit is 
evaluated by the standard extrapolation method. The numerical error after this extrapolation is estimated to be of order $10^{-3}$, 
so the gap is identified as the region of $\omega$ around $\omega \simeq 0 $ 
where the extrapolated $D(\omega)$ is less than $10^{-2}$.  
We also compute 
the ferrimagnetic order parameter per site  
\begin{eqnarray}
M&=& \sum_{a=1}^{3} ( \langle n_{a\uparrow } \rangle - \langle n_{a\downarrow } \rangle ) \nonumber
\end{eqnarray}
and the double occupancy per site 
\begin{eqnarray}
D_{occ.}= \frac{1}{3}\sum_{a=1}^{3} \langle n_{a\uparrow } n_{a\downarrow } \rangle = \frac{dE}{dU} \nonumber 
\end{eqnarray}
where $ \langle n_{a\sigma } \rangle $ and $\langle n_{a\uparrow } n_{a\downarrow } \rangle$ are the expectation values of $n_{a\sigma }$ and 
$n_{a\uparrow } n_{a\downarrow }$, respectively, with $a=$1, 2, and 3 being the sites in Fig.~\ref{fig:model}(a).

\begin{figure}
\includegraphics[width=0.43\textwidth,trim = 20 0 10 15,clip]{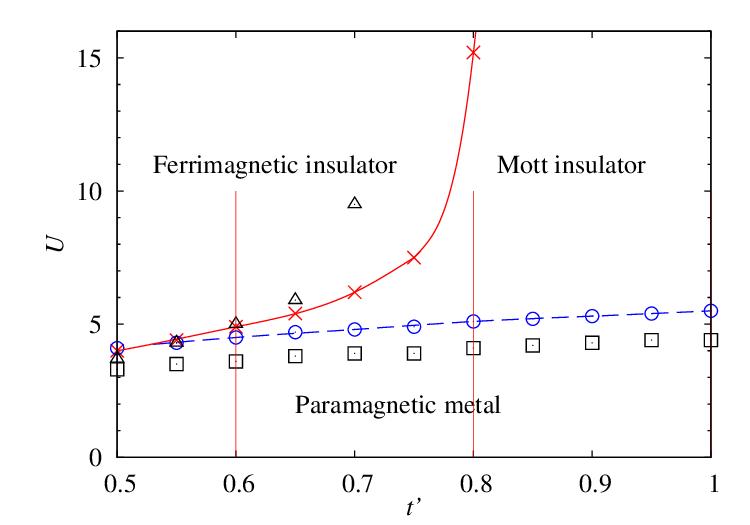}\\[-0.5em]
\caption{(Color online) Phase diagram of the Hubbard model on the anisotropic kagom\'e lattice 
at zero temperature and half-filling as a function of $t'$ and $U$ 
obtained by VCA, where the 12-site cluster is used (the crosses and circles). Lines are guides to the eye. 
The triangles and squares are the results obtained using the 6-site cluster. 
The crosses and triangles correspond to the ferrimagnetic and paramagnetic transition points and 
circles and squares are for the Mott transition points. 
\label{fig:phase-diagram}\\[-2.5em]}
\end{figure}

\begin{figure}
\includegraphics[width=0.42\textwidth,trim = 0 0 0 0, clip]{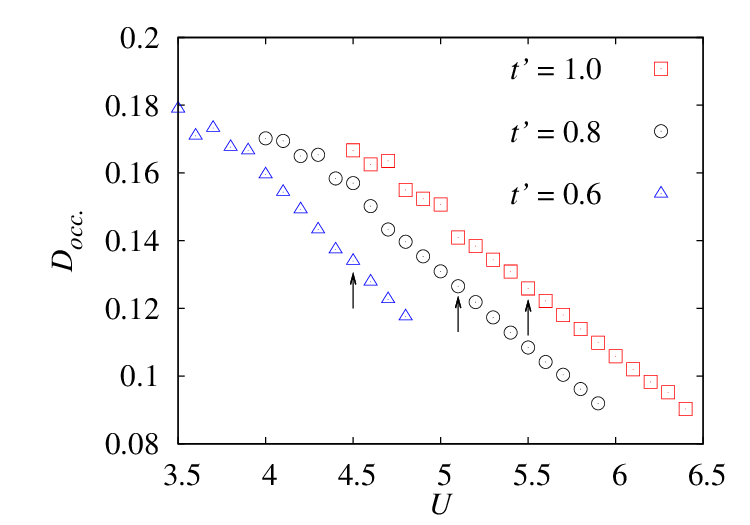}\\
\caption{(Color online) The double occupancy $D_{occ.}$ as a function of $U$ for $t' = 1.0$,  $0.8$, 
and $0.6$. The 12-site cluster is used. The three arrows indicate the Mott transition points. 
\label{fig:do}\\
}
\end{figure}

In Fig.~\ref{fig:phase-diagram}, we show the phase diagram at zero temperature and half-filling 
obtained from this analysis using the 12-site cluster. The results obtained using the 6-site cluster are also shown to 
quantitatively see the cluster size dependence. 
The critical interaction strength $U_{\rm F}$ separating the ferrimagnetic and MI phases 
rapidly decreases in the moderately frustrated region $t' = 0.5 \sim 0.7 $, showing that the ferrimagnetic fluctuations 
due to the relaxation of the geometric frustration is enhanced by the electron correlations. 
(At $t'=0.75$ $U_{\rm F}>20$ for the 6-site cluster.) 
In this region of $t'$, the ferrimagnetic phase is an insulator since there is a gap, and the transition 
between the ferrimagnetic and paramagnetic (including MI) phases is a level crossing (first order) because the 
ferrimagnetic solutions exist also $U < U_{\rm F}$ even though it is energetically disfavored there. 
The critical interaction strength $U_{{\rm MI}}$ separating the MI and metallic phases is slightly smaller  
than the noninteracting band width $W$, where $W = 6$ at $t'=1$ and $W=4\sqrt{2} \simeq 5.66$ at $t'=0$. 
$U_{{\rm MI}}$ decreases as the geometric frustration is relaxed, and the slope becomes steeper in moderately 
frustrated region. For the 12-site results, at $t'=0.5$, $U_{\rm F} = 4.0$ while $U_{{\rm MI}} = 4.1$ so the MI phase has disappeared. 
Taking into account the drastic growth of ferrimagnetic phase and the fact that $W$ remains almost the same, the decrease of $U_{{\rm MI}}$ 
according to the relaxation stems from the ferrimagnetic fluctuations. 
As for the Mott transition, we could not find out the Mott insulator and paramagnetic metal coexisting region of $U$ 
at half-filling within our two controlling parameters $\mu$ and $\mu'$. 
Also as will be shown later the Mott gap changes continuously as a function of $U$. 
Therefore, we could not find out an indication of the discontinuity at the Mott transition in this analysis. 
To supplement this analysis, we show in Fig.~\ref{fig:do} the double occupancy $D_{occ.}$ as a function of $U$ for the 12-site cluster, 
which also looks continuous at the transition point. 
In Ref.~\onlinecite{kawakami10} this transition is reported to be first order. 
First order Mott transitions are obtained in other models in the variational cluster approach with bath degrees of freedom and 
treating the hybridization between the bath sites and cluster sites as a variational parameter.\cite{Potthoff2,Potthoff3} 
In these analyses, the coexisting metal and insulator solutions, leading to the first order transition, differ by the value of these 
hybridization parameters, and these situations will be similar to the case of Ref.~\onlinecite{kawakami10}. 
Our analysis does not have bath degrees of freedom and technically this will be the origin of the difference. 
It remains to be clarified which is the correct picture. 

\begin{figure}
\includegraphics[width=0.47\textwidth,trim = 5 1 5 6,clip]{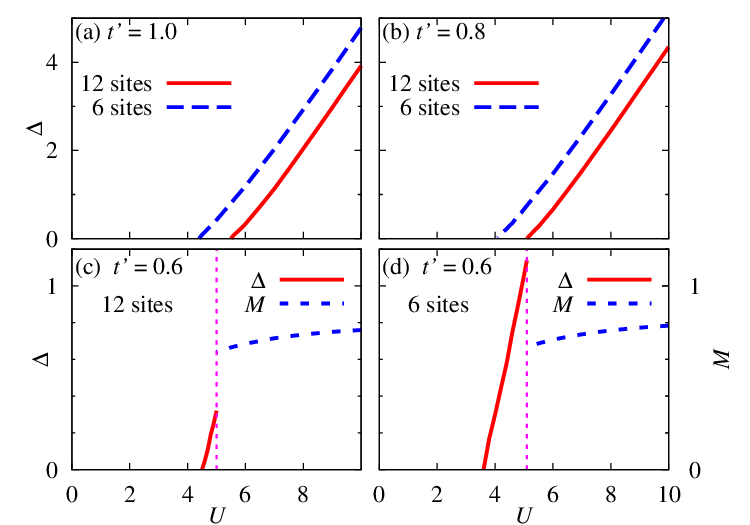}\\[-1.0em]
\caption{(Color online) Mott gap $\Delta$ as a function of $U$ at (a) $t' = 1.0$, (b) $t' = 0.8$, 
and (c), (d) $t' = 0.6$. These parameter regions correspond to the three vertical lines in Fig.~\ref{fig:phase-diagram}. 
For $t'=0.6$, order parameter $M$ is also included and the vertical lines separate the ferrimagnetic and MI phase. 
\label{fig:gap-af}\\[-2.2em]}
\end{figure}

Next we consider the cluster size dependence of our results. In general, $U_{{\rm MI}}$ is larger for larger clusters, 
since the kinetic energy of the cluster Hamiltonian can be larger for larger clusters. 
As for $U_{\rm F}$, when spin correlations are highly suppressed due to the frustration, 
cluster wave functions with small ferrimagnetic fluctuations play an important role to examine near 
the true minimum of the effective potential, so $U_{\rm F}$ is smaller for larger clusters. 
When the geometric frustration is moderate and spin correlations are not largely suppressed, 
the difference of cluster kinetic energies due to the cluster size 
becomes more important to determine the phase boundary, so $U_{\rm F}$ becomes larger for larger clusters. 
Our result is consistent with this general argument on the cluster size dependence. 
Quantitatively, $U_{\rm F}$ is almost the same for the 12- and 6-site clusters at $t'=0.6$, and $U_{\rm F}$ is smaller for 
the 12-site clusters for $t' > 0.6$. Relatively large difference of $U_{\rm F}$ between the 12- and 6-site cluster results 
in strongly frustrated region indicates strong suppression of the spin correlations. 
The difference of $U_{{\rm MI}}$ between the 12-site and 6-site analysis is less than 20\% of $W$. 
The behavior of our $U_{{\rm MI}}$ according to the relaxation of the frustration is qualitatively consistent with the 
previous results\cite{kawakami10}, though our values for $U_{{\rm MI}}$ are relatively small compared to those 
in Ref.~\onlinecite{kawakami10}. At present the origin of these discrepancies are not clear to us.

%
%

\begin{figure}
\subfigure{\includegraphics[width=0.47\textwidth, trim = 100 150 250 740, clip]
 {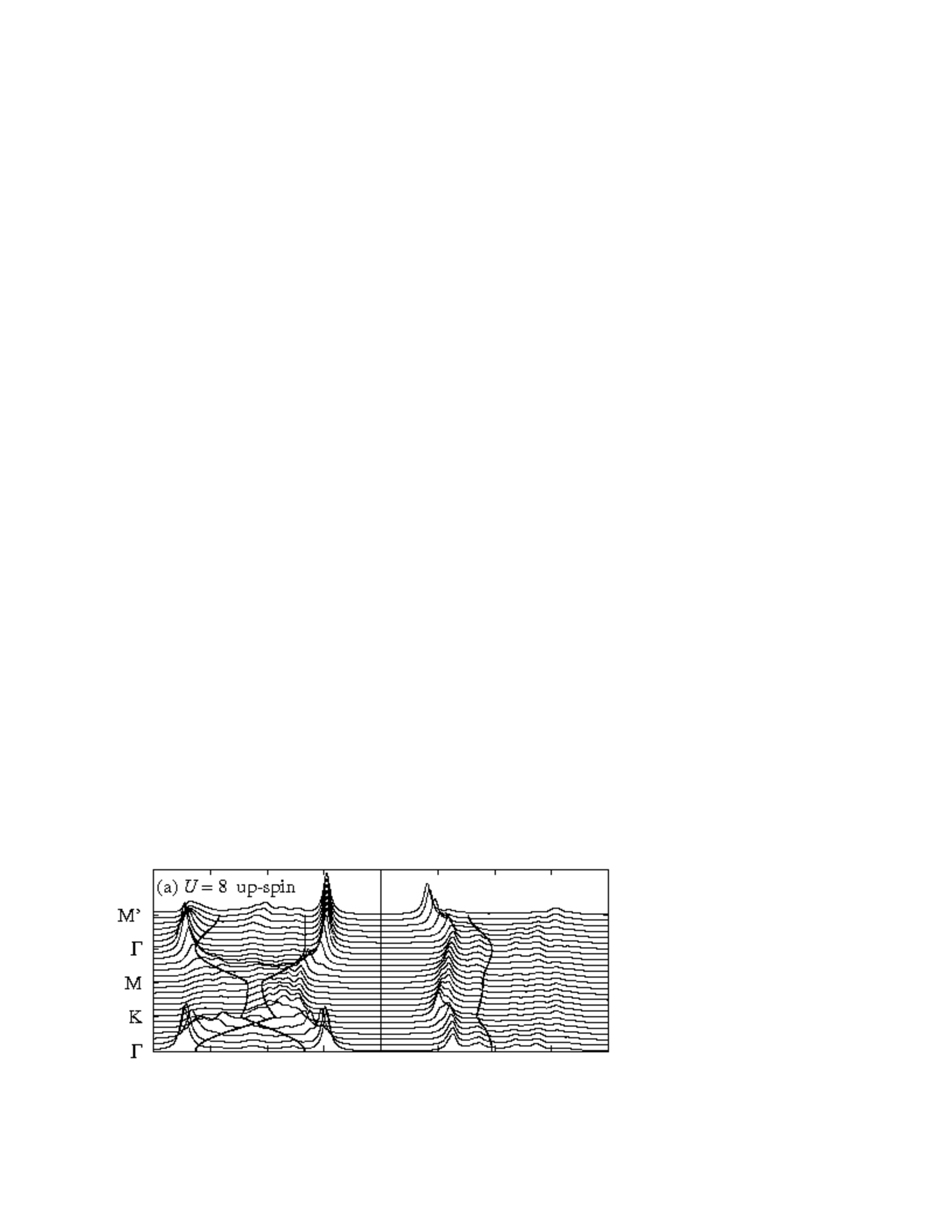}\label{fig:sp-075-u8up}}\\[-1.6em]
\subfigure{\includegraphics[width=0.47\textwidth, trim = 100 150 250 740, clip]
 {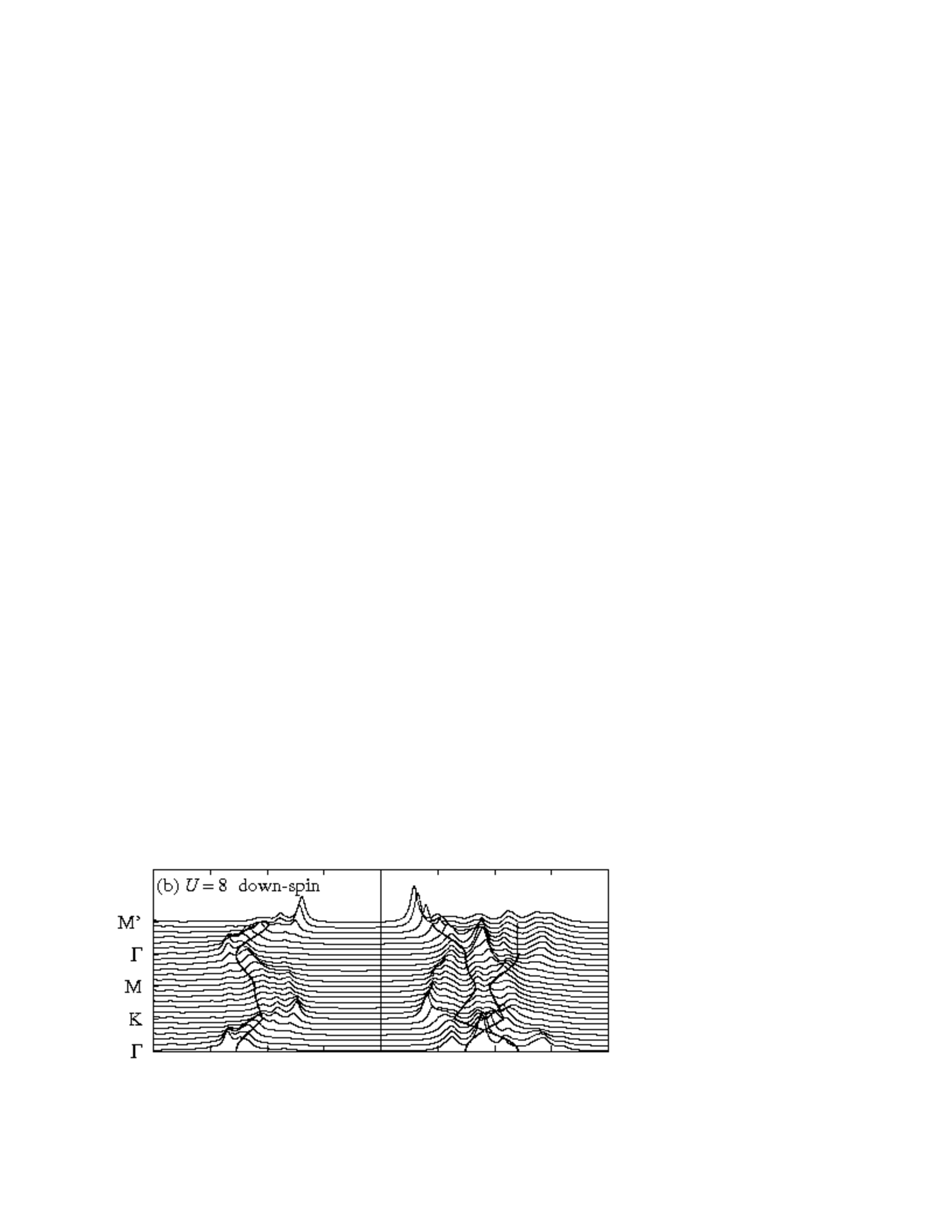}\label{fig:sp-075-u8dw}}\\[-1.6em]
\subfigure{\includegraphics[width=0.47\textwidth, trim = 100 150 250 740, clip]
 {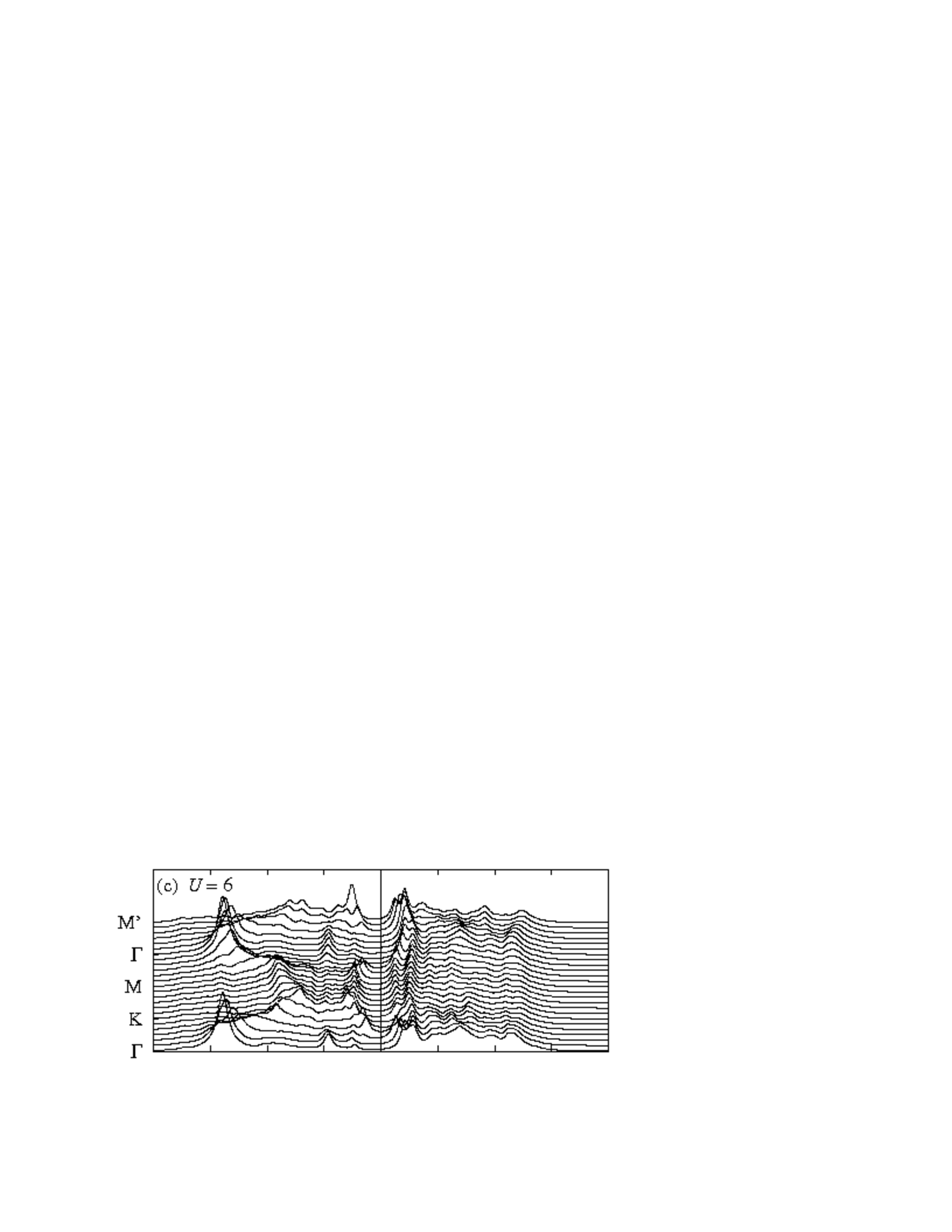}\label{fig:sp-075-u6}}\\[-1.75em]
\subfigure{\includegraphics[width=0.47\textwidth, trim = 100 120 250 720, clip]
 {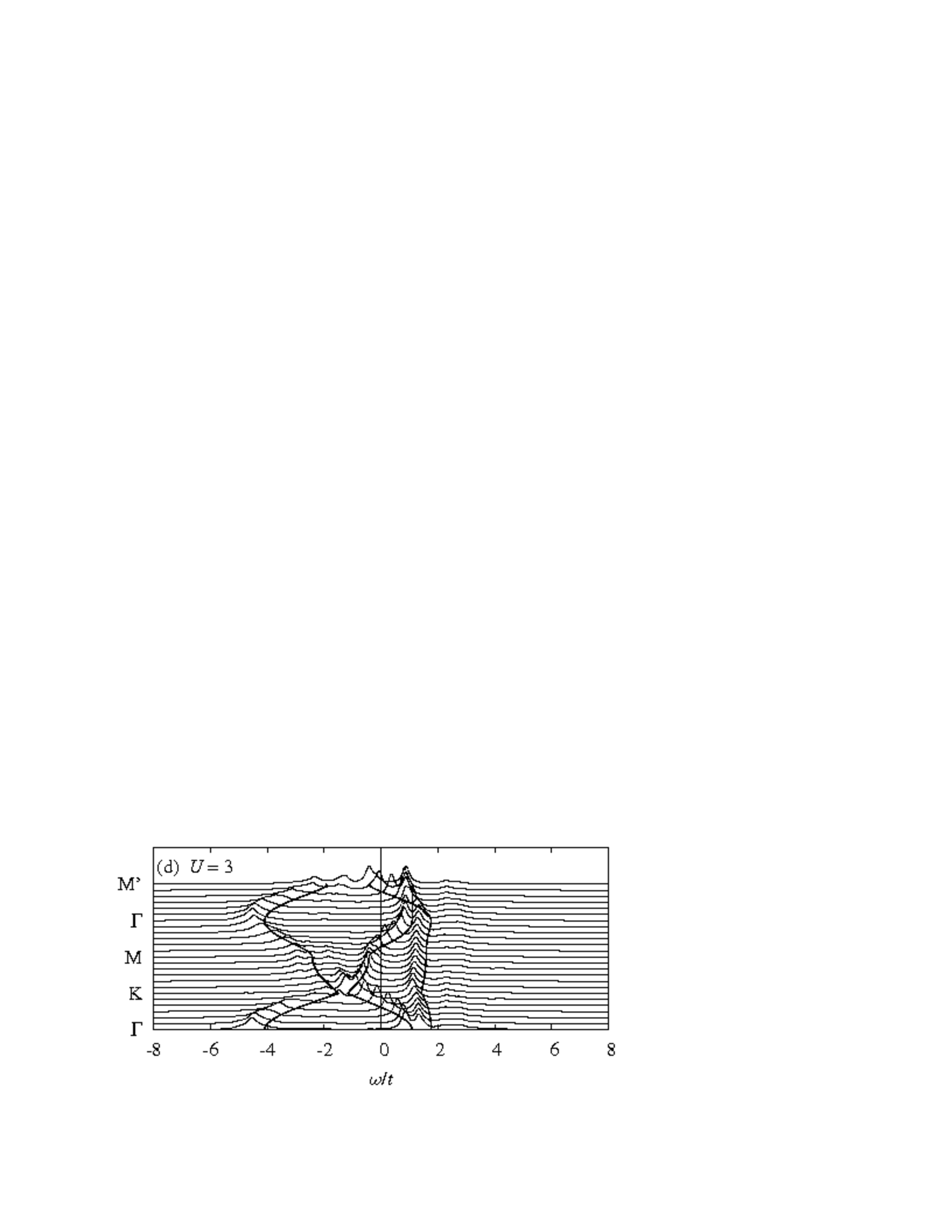}\label{fig:sp-075-u3}}\\[-1.5em]
\caption{Spectral density at $t' = 0.75$ 
for (a), (b) $U=8$ (ferrimagnetic state), (c) $U= 6$ (MI state), and (d) $U=3$ (metallic state) along the dotted line in Fig. \ref{fig:model}(b). 
The Lorentzian broadening with $\eta = 0.15t$ is used in all the cases. 
In (a) and (b), the solid lines are the mean-field SDW dispersion for the same values of $U$, $t'$ and $\mu$. 
In (d), the solid lines are the noninteracting band structure. In (a), (b), and (c) the peaks are scaled by $5$ compared to (d). 
\label{fig:sp}\\[-2.0em]}
\end{figure}   

\begin{figure}
\includegraphics[width=0.47\textwidth,trim = 0 0 0 5,clip]{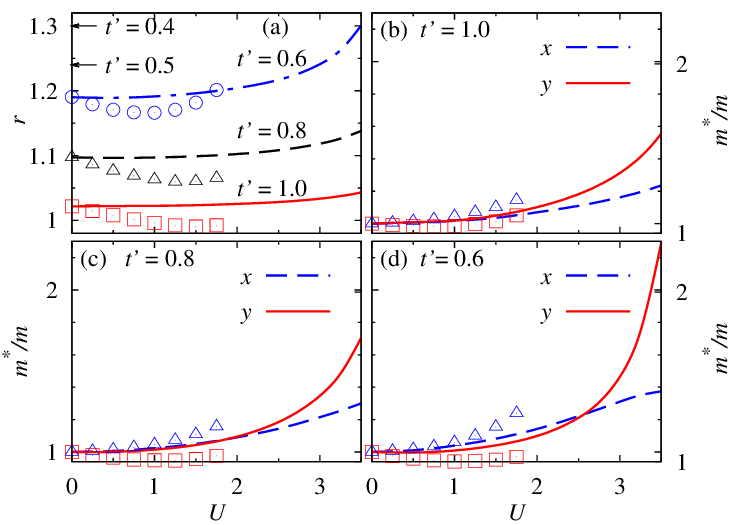}\\[-0.0em]
\caption{(Color online)  (a) Ratio $r$ as a function of $U$ for $t'=1.0$, $0.8$, and $0.6$. 
The two arrows indicate the value of $r$ for noninteracting band at $t'=0.5$ ($r=1.24$) and at $t'=0.4$ ($r=1.30$). 
(b)$\sim$(d) Mass enhancement factor $m^{*}/m$ in the $x$ and $y$ directions as functions of $U$ for (b) $t'=1.0$, (c) $t' = 0.8$, and 
(d) $t' = 0.6$. The lines are obtained using the 12-site cluster and symbols (squares, triangles, and circles) are the 
results with the 6-site cluster. 
In (a) the squares, triangles, and circles correspond to $t'=1.0$, $0.8$, and $0.6$, respectively. 
In (b)$\sim$(d) the triangles correspond to the $x$ direction while the squares correspond to the $y$ direction.   
\label{fig:mass}\\[-2.5em]}
\end{figure}

In Fig.~\ref{fig:gap-af} we show the Mott gap $\Delta$ and ferrimagnetic order parameter $M$ as functions of $U$ 
for $t' = 1.0$, $0.8$, and $0.6$ corresponding to the three vertical lines in Fig.~\ref{fig:phase-diagram}. 
$\Delta$ monotonically decreases as $U$ decreases in all cases and $M$ is always smaller for the 
12-site cluster at $t' = 0.6$. The gap $\Delta_F$ in the ferrimagnetic phase is slightly larger than $\Delta$ if $U$ is the same. 
For example, for $U=10$, $\Delta=4.35$ at $t'=0.8$ while $\Delta_F=4.91$ at $t'=0.7$ and $\Delta_F=5.54$ at $t'=0.5$ 
in the 12-site analysis. 

%
%

In Fig.~\ref{fig:sp} we show the spectral weight function $\rho(\omega, k)$ calculated using the 12-site cluster 
for solutions corresponding to (a), (b) the ferrimagnetic phase (up and down spin parts are plotted separately), 
(c) the MI phase, and (d) the metallic phase at $t' = 0.75$. 
In (a) and (b), the mean-field spin-density-wave (SDW) dispersion is also included (solid lines) to see its general features, 
where $M=0.92$ in the mean-field solution while $M=0.72$ in VCA. 
In (d), the noninteracting band structure is also plotted (solid lines) for comparison. 
In (a) and (b), the dispersion is largely affected due to the electron correlations compared to the mean-field solution. 
In (c), the SDW dispersion disappears and the spectral function displays a Mott gap across all wave vectors. 
The gap is smaller compared to (a) and (b) since $U$ is smaller. Comparing with (d), in (c) it looks that 
the lowest energy band in (d) is shifted downwards and the second band splits into lower and upper Hubbard bands while 
the top flat band remains almost the same. 
In (d), we notice that the spectral function is consistent with the Fermi liquid state and the interacting bands slightly shrink toward the Fermi surface, 
leading to heavy fermion behavior.

To study it in detail, 
we consider well below the MI transition line where Fermi liquid natures are confirmed from the behavior of the 
spectral function and compute the mass enhancement factor $m^{*}/m$ along the $x$ and $y$ directions in $k$ space, 
where the $x$ direction corresponds to the direction of $t'$ hopping in real space. 
Near the Fermi surface the position of the peak $(\omega, k) = (\omega_F + \delta \omega, k_F + \delta k) $ of $\rho(\omega, k)$ 
changes according to the relation $|\delta \omega | = (k_F/m^{*}) \delta k $ and the band mass $m^{*}$ is calculated 
using this relation. 
We also compute the ratio of the Fermi momenta in the $x$ and $y$ directions, $r = k_{yF} / k_{xF}$. 
As $t'$ decreases, the noninteracting Fermi surface slightly shrinks in the $x$ direction and slightly 
evolves in the $y$ direction\cite{imai03}, 
so $r$ is a measure of the anisotropy including the effect of the electron correlations. 
Even though precise values of these quantities may depend on the lattice geometry, 
we show in Fig. \ref{fig:mass} $r$ and $m^{*}/m$ in the 
$x$ and $y$ directions as functions of $U$ for 
$t' = 1.0$, $0.8$, and $0.6$ in our lattice geometry, to see general features about 
the effect of the electron correlations on these quantities. The lines are the results obtained using the 12-site cluster and 
symbols are the results with the 6-site cluster. 
At $t'=0.6$, $r$ rapidly grows around $U \simeq 3$ for the 12-site cluster. This tendency is also observed 
for the 6-site cluster, where $r$ turns to grow around $U \simeq 1.5$. 
The rapid growth of $r$ indicates that the effective anisotropy is enhanced due to the electron correlations in moderately frustrated region. 
In fact, for the 12-site cluster, the value of $r$ at $t'=0.6$ and $U=3.5$ is equal to that of noninteracting band at $t'=0.4$. 
The analysis of the effective anisotropy was also done in Ref.~\onlinecite{kawakami10} by considering the renormalization of the 
hoping parameters and our results are qualitatively consistent with their analysis. 
This enhancement of effective anisotropy indicates that the spin fluctuations due to the relaxation of the 
geometric frustration are also enhanced by the electron correlations. 
Within our analysis, this indication is consistent with the rapid growth of the ferrimagnetic phase above $U_{MI}$, 
and it is demonstrated by the analysis of the spin correlations.\cite{kawakami10} 
As is shown in Fig.~\ref{fig:mass}(b)$\sim$(d), the heavy fermion behavior is observed in all cases for the 12-site cluster. 
For the 6-site cluster sizable mass enhancements are not observed and the 6-site cluster may not be large enough for subtle analysis 
related to the spectral function. 
The growth of $r$ affects also $m^{*}/m$ since $k_F$ enters into the calculation of $m^{*}$. 
In general, $m^{*}$ is enhanced in the $y$ direction and suppressed in the $x$ direction since the Fermi surface 
shrinks in the $x$ direction and evolves in the $y$ direction. This appears largely in moderately frustrated region 
due to the rapid growth of $r$, as is observed around $U \sim 3.5$ for the 12-site cluster results in Fig. \ref{fig:mass}(d). 
Therefore the anisotropy of the effective masses is enhanced in moderately frustrated region. 

In summary  we have investigated the ferrimagnetism and Mott transition on the anisotropic 
the kagom\'e lattice using VCA. 
The phase diagram at zero temperature and half-filling is determined. 
The ferrimagnetic phase rapidly grows in moderately frustrated region and the MI phase disappears there. 
In the metallic phase, heavy fermion behavior is studied and the mass enhancement is computed. 
Enhancement of spatial anisotropy due to the electron correlations is also observed for moderately frustrated 
region and its effect on the heavy fermion behavior is discussed. 
Thus, the interplay between the spin correlations and Mott transition is quantitatively studied above and below the 
metal-insulator transition. 

One of us (A.Y.) would like to thank N.~Fukui, K.~Kurasawa, H.~Mikami, and H.~Nakada for useful discussions on numerical analysis. 
This work was supported in part by Kakenhi Grant No.~22540363 of Japan. A part of computations was done at 
Research Center for Computational Science, Okazaki, Japan.

\end{document}